# Evaluating CBR Similarity Functions for BAM Switching in Networks with Dynamic Traffic Profile

Eliseu M. Oliveira
Salvador University, Salvador, Brazil
eliseu@gmail.com

Rafael Freitas Reale
IFBA, Bahia, Brazil
reale@ifba.edu.br

Joberto S. B. Martins
Salvador University, Salvador, Brazil
joberto.martins@unifacs.br

***Abstract*—In an increasingly complex scenario for network management, a solution that allows configuration in more autonomous way with less intervention of the network manager is expected. This paper presents an evaluation of similarity functions that are necessary in the context of using a learning strategy for finding solutions. The learning approach considered is based on Case-Based Reasoning (CBR) and is applied to a network scenario where different Bandwidth Allocation Models (BAMs) behaviors are used and must be eventually switched looking for the best possible network operation. In this context, it is required to identify and configure an adequate similarity function that will be used in the learning process to recover similar solutions previously considered. This paper introduces the similarity functions, explains the relevant aspects of the learning process in which the similarity function plays a role and, finally, presents a proof of concept for a specific similarity function adopted. Results show that the similarity function was capable to get similar results from the existing use case database. As such, the use of similarity functions with CBR technique has proved to be potentially satisfactory for supporting BAM switching decisions mostly driven by the dynamics of input traffic profile.**

## I. Introduction

Most of the actual and frequently used network applications commonly require a multiservice network with bandwidth availability and do require support for distributed high performance processing. That is, for instance, the case of high definition video and image transmission, remote instrumentation and analysis, and parallel grid processing, among many others [1]. Another characteristic that actual networks have inherently to comply is with the fact that, in real scenarios, bandwidth may become scarce and that the input traffic profile for actual networks is highly variable.

One fundamental problem for the drafted scenario is how to provide resources (bandwidth, computer power, storage, other) in such a way that applications have enough network quality to execute and provide good user experience. The basic solution for this problem consists then in finding and managing the best possible configuration for the network for various parameters, bandwidth being one of the most important ones and the focus of this paper.

The Bandwidth Allocation Models (BAMs) over IP/MPLS networks as discussed in [2] [3] are the solution used to arbitrate bandwidth for the dynamic input traffic in the context of the similarity evaluation and learning approach discussed in this paper. In brief, input traffic demands resource (bandwidth) for the BAM model (behavior) in use and the requested resource may be granted or blocked and currently setup LSPs (Label Switched Paths) might be preempted or teardown to return resource (bandwidth) for the BAM [4].

Another important consideration in terms of the similarity function evaluation scenario is that there are various BAM models. Each BAM model implements in fact a distinc network behavior depending on the input traffic. As such, the essence of the management problem discussed here is what is the best possible BAM model (behavior) to use and when should the management system change (switch) to them. This is the scenario where the management should learn and where the similarity evaluation in the learning context applies.

The learning approach adopted for the BAM switching scenario presented with an IP/MPLS network infrastructure is the Case-Based Reasoning solution [5].

In brief, this article presents an evaluation of similarity functions alternatives that are necessary for the implementation of a Case-Based Reasoning learning approach. CBR will gradually learn how to switch between models and/or reconfigure actual BAMs. This BAM switching and/or reconfiguration are the way to react to the dynamicity of the IP input traffic profiles and to meet the network politicies defined by the network manager.

## II. CBR – Case-Based Reasoning

Case-Based Reasoning (CBR) is, in brief, a method that supports the solution of new problems in any area using the experience acquired in previous cases. CBR functions as a cognitive model that allows to mimic human to solve real problems based on learning from similar past case solutions [5].

Concrete example of cases is the major technique used by CBR in solving a new specific case. CBR applications provide the solution to a problem and the ability to deduce and justify decisions and actions taken by using the experience obtained in solving previous similar cases.

The basic elements of a CBR system are the representation of knowledge, the measure of similarity, the adaptation capability and learning [5]. For knowledge representation are used case descriptions and the concrete experiences obtained with their solution that is stored in the CBR database. In terms of the similarity, a database search looks for relevant cases that are like the case we want to solve. Although the similar case would not be exactly the same, CBR adapts the solution found that presents characteristics of the current case. Learning takes place when the case-base system is continually updated whenever the solution is satisfactory.

The CBR more accepted model for the CBR Cycle is called 4R and includes the following phases [5] (Figure 1):

- Retrieve similar cases from the database;
- Reuse these cases for solving the problem;
- Review the proposed solution; and
- Retain the experience by storing it in the CBR database for reuse.

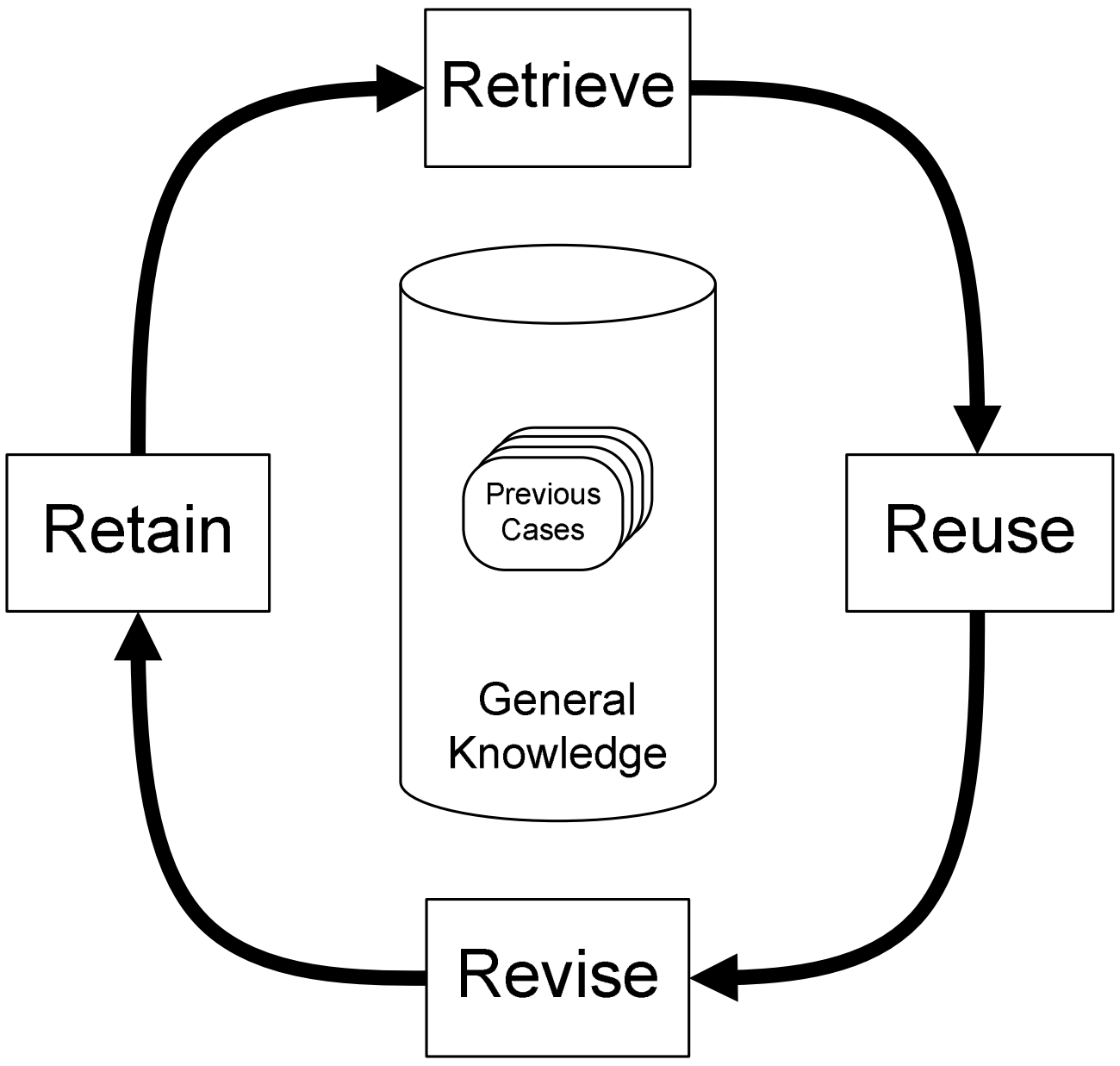

Figure 1 – CBR cycle

The following section presents the proposal for a CBR-BAM Module supporting the switching of Bandwidth Allocation Models (BAM) in the context of an IP/MPLS (MultiProtocol Label Switching) network subject to dynamic input traffic. The basic idea is that the CBR-BAM module could assist the process of choosing a new BAM model (behavior) that suits better any new input traffic condition (input traffic dynamics).

## III. CBR-BAM Module Proposal

### *A. Problem Modeling*

The first step in the process of using CBR in any area of knowledge is the domain definition for the problem.

In the context of IP/MPLS networks, this work aims to find with autonomic characteristics what is the best BAM model and/or configuration to troubleshoot specific network problems such as: low bandwidth utilization in a BAM TC (Traffic Class), high number of LSP (Label Switched Paths) request blocking on a BAM TC or high preemption rates and bandwidth devolution by link or TCs, among others.

Once defined the domain, it should be defined with great accuracy the attributes that will be used as “indices” for searching similar cases. These indices are variables defined by a n-tuple key/ values that will be used to recover similar cases at the CBR database.

The similar case CBR database search requires a similarity computation with four components:

- Context information: information acquired before the recommendation process that identify the context in which the similarity should act;
- CBR-BAM Problem: symptoms and alerts that characterize the current problem;
- Measurements: information obtained from network analysis and monitoring;
- Similarity algorithm: the algorithm that realizes the comparison between the previously stored CBR case and input case in actual CBR cycle.

### *B. Modeling Case*

In the modeling case discussed in this paper, the indices will be created using the context attributes and network monitoring parameters. The context attributes are derived from the atual network configuration and policies defined by the manager. The CBR-BAM problem will trigger the process and works as a kind of cathalizer in searching similar solutions.

The contextual attributes proposed are as follows:

- BAM model (behavior) being used;
- Network operational parameters “*Limits*”:
  - Number of *preemptions* (range), maximum LSP *blocking* rate, other parameters limits.
  - *Limits* are defined by the network manager and apply either to TCs or links.
- Bandwidth constraints associated with the link (BCs).

Any change on these defined attributes leads to a case of learning in a new associated context.

The CBR-BAM problem corresponds to the symptoms and alerts that caractherize the problem to which a similar solution is searched. The symptoms, as an example, are emergencial alerts or periodic diagnostics received by the manager that are not in accordance with acceptable values previously defined by the manager (the "limits").

The *measurements*, toghether with the context information and CBR-BAM problem, provide a snapshot of the network operational parameters (links, other). These measurements are essential to reflect network profile alteration and the following measurements were adopted in the CBR-BAM module proposal:

- *Blocking* rate, *Utilization*, *Preemption* and *Devolution* per TC.

### C. Similarity Functions

After defining the indexes, it is necessary to define the search methodology and the similarity function adopted for comparing similar cases. There are many similarity functions, each one with a specific purpose.

The similarity of one case "T" in relation to others available cases (S, Z, K, …) is defined as as a mean among distinct similarities for parts of the case. As an example, let´s consider the case "T" with three indexed attributes ($T_j$, $T_i$ and $T_m$). This case will be similar to (like) another case "S" if the indexed attributes ($S_j$, $S_i$ and $S_m$) were similar to ($T_j$, $T_i$ and $T_m$) among themselves.

This partial similarity, by attribute, is called local similarity. There are various functions for calculating local similarity, and the choice depends on the type of the attribute used.

The most elementary local similarity function assumes that if an attribute $T_i$ is equal to the other attribute $S_i$ the similarity between them is equal to 1, otherwise it is null, as indicated in the function below.

$$f\ (T_i, S_i) = \begin{cases} 1, if\ T_i = S_i \\ 0, if\ T_i \neq S_i \end{cases}$$

This function is the one used to calculate the similarity for BAMs used in cases and uses the contextual information attribute of this proposal. In this specific similarity case a full match is required, or effectively, we have a different approach or contex in use.

However, this type of similarity function can not be applied to all *indices* because it does not consider approximate values for the *indice* and, as such, is not indicate for *indices* based on numerical values.

For the remaining of the similarity function discussion we consider that:

- T and S are respectively the input and stored cases;
- "i" is an attribute; and
- "f" is the local similarity function for "i" attribute.

To calculate the similarity between numerical attributes, it can be used the distance between them, represented by the module of the difference between them as indicated in equation 1 [5].

$$f(T_i, S_i) = |S_i - T_i| \qquad \text{Equation 1}$$

The linear function (Equation 2) is another option to calculate the similarity between numerical atributes. In this case, the similarity increases for smaller distances but the range of variation of the numbers is considered.

$$f(T_i, S_i) = \frac{1-|S_i - T_i|}{(max - min)} \qquad \text{Equation2}$$

In this proposal, the linear function will be used for the numeric attributes *blocking*, *preemption* and *devolution*. For example, given two cases (S, T) where *blocking* for $TC0_i$ ($T_i$) is 80% and for $TC0_i$ ($S_i$) is 70%, the local similarity will be 90%.

An intermediate option between the basic local similarity function and the linear function is the ladder function (Equation 3). In this case, a maximum "k" value for the difference between the two values is defined. If the actual difference between the attributes is less than or equal the absolute value "k" the similarity will be 1, if not 0.

$$f\ (T_i, S_i) = \begin{cases} 1, if\ |T_i - S_i| \leq k \\ 0, if\ |T_i - S_i| > k \end{cases} \qquad \text{Equation 3}$$

The ladder similarity function is used in the CBR-BAM module for the bandwidth attribute. As an example, the values in Table 1 are considered all 100% similar, if the value of "k" is 256. If "k" is 128, only the lines in bold will be considered similar.

Table 1 –Bandwidth similarity with the ladder function

| BC0*i* | BC0*j* |
|---|---|
| **250** | **200** |
| 256 | 512 |
| 256 | 500 |
| **512** | **640** |

The local similarities are the base to calculate and define the global similarity. In effect, the global similarity is defined as a computed mean of the distinct local similarities for the attributes. Like local similarity, the global similarity has several calculation approaches and methods being at the discretion of the solution developer find out the function that best suits his domain.

The CBR-BAM module, focus of this paper, is used in a IP MPLS network and works mostly with quantitative variables. As such, we chose to use the closest neighbor algorithm, since it returns similarities using only a statistical model.

The CBR-BAM module requires that cases belonging to a context should not return similar cases belonging to another context. That is so, because the requirements and needs of each context might be different. For example, possibly a network state will be recommended for a management policy, but this same state will not be appropriate for other management policy.

The Nearest-Neighbor algorithm (Equation 4) is the option used for the global similarity function calculation within the CBR-BAM module being described [6]. The Nearest-Neighbor algorithm calculates the sum of the similarities of the "n" indexed attributes. A variation of this algorithm considers the weight of each attribute to determine the similarity.

$$Sim\,(T,S) = \frac{\sum_{i=1}^{n} f(Ti, Si) \times Wi}{\sum_{i=1}^{n} Wi} \qquad \text{Equation 4}$$

Where:

- "T" is the input case
- "S" is the case stored in CBR database
- "n" is the attribute number of each case
- "i" is an individual attribute
- "f" is the local similarity function for attribute "i" in cases T and S
- "W" is the weight given to the attribute "i"

### D. *Similarity Levels*

In the CBR-BAM Module the global similarity function is calculated with three hierarchical levels.

At the first level, we have four variables: the current BAM in use, the bandwidth used, the current network metrics and the network manager defined tolerance for these metrics. Since the BAM variable represents the BAM model in use and has no variance (unlike numerical parameters that do have), its value is immediately assigned to the global similarity function.

At the second level, we consider the variance of the variables *tolerance*, *bandwidth* and *measurements*. The tolerance variable is applied to the *preemption*, *devolution*, and *blocking* variables. The *bandwidth* variable returns the attributes with values corresponding to the bandwidth limits on each BC (Bandwidth Constraint) for the case. The *measurement* variable contains the network monitored parameters such as *utilization*, *preemption*, *blocking* and *devolution*.

At the third level, there are final derivations of the tolerance and measurement variables. At this level, we obtain the actual values of the metrics measured for the nTCs. Tolerances are implemented as the values that the network manager assigns as acceptable for each TC for preemption, devolution and blocking. Measurements are the actual network values for blocking, preemption and return by TC. The se values are measured in the network and monitored at the moment of similarity calculation. The three levels are shown in figure 2

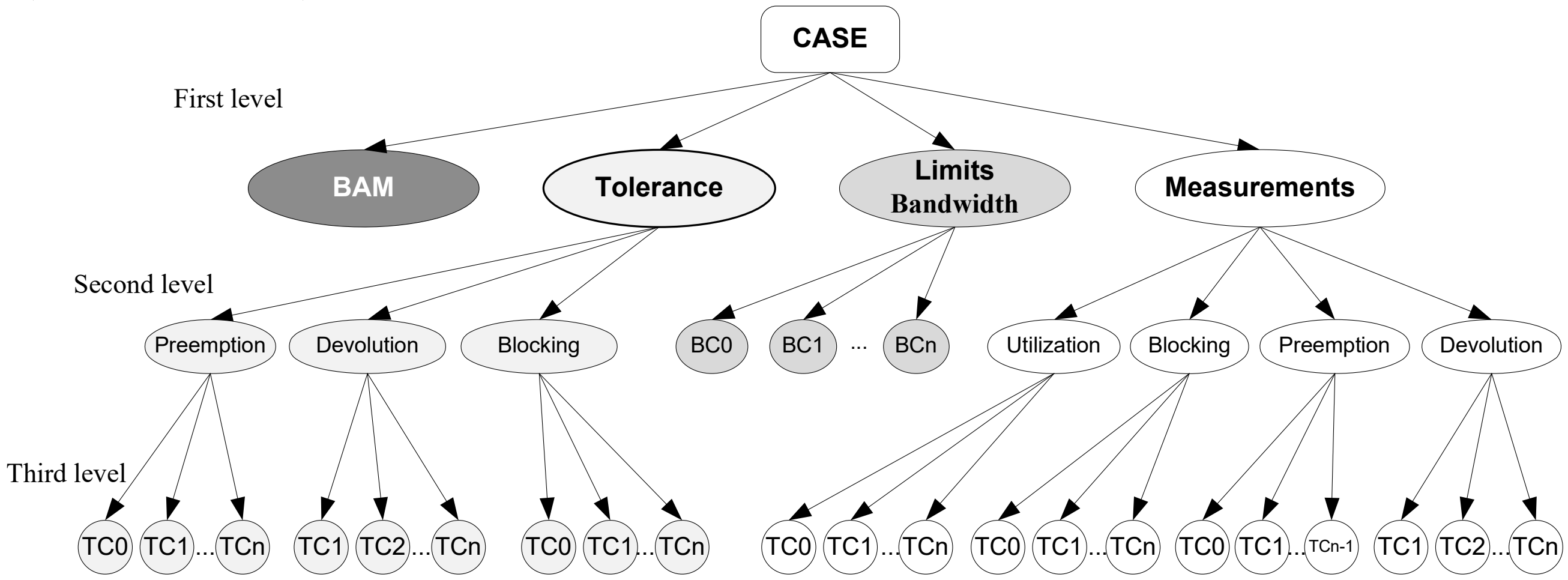

Figure 2 – Similarity Levels

## IV. THE CBR-BAM MODULE EXECUTION

The CBR-BAM module is based on CBR4R cycle. The four phases of the CBR cycle will be presented to illustrate the whole cycle and highlight the similarity function role in the entire process.

The CBR cyclestarts with the CBR-BAM module receivingan "alert" indicating that the limits previously set by the network manager are not being obeyed. At that moment, the CBR-BAM modulegets from the network all values for the indexed attributes containing the characteristics that describe the current state of the network. This set of information is named "Current Case" or "case" for short. Having the "case" the CBR-BAM module triggers the first step of the CBR 4R cycle that as shown in Figure 01.

### A. *First phase: Recovery*

At this stage, the CBR-BAM module checks the existence or not of a stored case with similar characteristics in relation to the actual case that generated the alarm. This phase returns the most significant cases (similar ones) towards a possible solution for the current case. If a similar case is found, the next phase (reuse) is called.

If there is no similar case to the current case, the network manager is notified and he must manually provide a solution to the current problem. The current case, along with the manager's solution are compiled and becomea "New Case" that issent for execution and then checked for retention.

Alternatively, the CBR cycle can execute without the intervention of the manager. With this option, an arbitrary solution is attributed to the current problem even without a match with the case base. In case the attributed solution is satisfactory, the case is stored in the CBR database, otherwise, a new solution is proposed and the previous alternative is stored on the negative cases database (not to be used in the future). This process is repeated until we have an adequate solution.

### B. *Second phase: Reuse*

Once a found case (positive case database) is considered similar (similar case) to the current case, the reuse phase composes the existing solution with the similar case to create a “new case”.

New Case = current case + similar solution

This new case is then compared with the database of negative cases. If it is not found in the negative database, the “new case” is considered ready for the next stage and is then forwarded for review. Otherwise, it is discarded and the initial phase is started again and the second most similar case is selected to repeat the process.

### C. *Third phase: Review*

In their review phase, the “new case” is previously tested to see if any adaptation is required for the solution before the “new case”is send to execution. The revision can be done in two ways: automatically, being performed by a simulated test or; manually, with manager intervention. It is also possible that a case does not need to be revised being used as such. In this case, the new case is send to execution to solve the problem.

### D. *Fourth stage: Retention*

The retention main objective is learning. So, the last phase of the CBR cycle verifies if the solution adopted is effective. If the solution is satisfactory, the new case is stored on the database of positive cases. If for some reason the solution adopted in the case is not satisfactory, it is then stored in the negative database base and the CBR 4R cycle starts again. The cycle only ends when a solution is accepted as adequate and a new case is stored on the database of positive cases.

## V. CBR-BAM Module Simulation

To validate the similarity function approach of the CBR-BAM Module, a proof of concept was constructed to simulate a scenario considering all previously defined attributes and that all indexed attributes have the same weight.

Three user profiles (corresponding to the tolerable values of *blocking*, *preemption* and *devolution* in each TC) are created for users Carlos, Marcos and Lucas as indicated in Table 2.

Table 2 –User profile - Tolerance

| User Name | Blocking | | | Preemption | | | Devolution | | |
|---|---|---|---|---|---|---|---|---|---|
| | TC0 | TC1 | TC2 | TC0 | TC1 | TC2 | TC0 | TC1 | TC2 |
| Carlos | 70 | 65 | 60 | 80 | 70 | 0 | 0 | 70 | 80 |
| Marcos | 60 | 50 | 40 | 90 | 80 | 0 | 0 | 80 | 90 |
| Lucas | 65 | 60 | 70 | 70 | 60 | 0 | 0 | 60 | 70 |

It is important to note that for this proof of concept it is understood that network user profiles do not change over time. As observed in Table 2, the use of three traffic classes TC0, TC1 and TC2 is defined for the BAM in use.

Two groups of values were used for the *bandwidth* variable and these values were combined with each user profile generating a total of six possible cases for the base. The values used are indicated in Table 3.

Table 3 – Bandwidth per BCs

| Bandwidth (Mbps) | | |
|---|---|---|
| BC0 | BC1 | BC2 |
| 256 | 512 | 1024 |
| 128 | 256 | 512 |

The values of *tolerance* and *measurement* of the CTs were filled so that each case at the CBR database represents only one case. For example, if a case presents a problem of low utilization, only the *utization* variables for TC0, TC1 and TC2 present values below those tolerated by the network manager. The possible values for utilization, blocking, preemption and devolution per CTs are expressed in Tables 4. The values are indexed in relation to the BAM model (behavior) options considered in the simulation:

- MAM – Maximum Allocation Model [7]
- RDM – Russian Dolls Model [8]; and
- ATC-S – AllocTC-Sharing [9].

Table 4 – Maximum and minimum values for utilization and blocking

| | Utilization | | | Blocking | | |
|---|---|---|---|---|---|---|
| BAM | TC0 | TC1 | TC2 | TC0 | TC1 | TC2 |
| MAM | 0-100 | 0-100 | 0-100 | 0-100 | 0-100 | 0-100 |
| RDM | 0-100 | 0-100 | 0-100 | 0-100 | 0-100 | 0-100 |
| ATC-S | 0-100 | 0-100 | 0-100 | 0-100 | 0-100 | 0-100 |

Table 5 – Maximum and minimum values for preemption and devolution

| | Preemption | | | Devolution | | |
|---|---|---|---|---|---|---|
| BAM | TC0 | TC1 | TC2 | TC0 | TC1 | TC2 |
| MAM | 0 | 0 | 0 | 0 | 0 | 0 |
| RDM | 0-100 | 0-100 | 0 | 0 | 0 | 0 |
| ATC-S | 0-100 | 0-100 | 0 | 0 | 0-100 | 0-100 |

Considering that, depending on the BAM used, there is no value for some TCs and combining the valid possibilities of TCs with the other six possibilities previously discussed and the three types of BAM used in this proof of concept we obtained a total of 54 cases in the CBR database.

Considering now that each case in the database is composed by a set of characteristics that describes the problem and a solution to the presented, the cases of the initial base were solved using a set of criterias presented in Table 4. For example, if a case presents a high preemption problem and this case has the RDM as the initial BAM, it is expected that the CBR recommendation suggests as a solution the change to the MAM.

Table 6 – Recommended solution for each problem by BAM

| | MAM | RDM | ATC-S |
|---|---|---|---|
| Low Use | ATC-S | ATC-S | ATC-S |
| HighLock | MAM | MAM | MAM |
| High Preemption | - | MAM | MAM |
| High Devolution | - | - | RDM |

To execute the simulation runs, a new situation was created where a fourth user (Fred) with distinct requirements from other managers and using MAM as the BAM model had a low utilization problem in TCs. The test succeeded in returning as a solution another case that recommended the use of MAM as a solution. Other scenarios were created and the same results were obtained as summarized in Table 5 with a satisfactory set of recommendationsfor the first round of the CBR cycle.

Table 7 – Case basis ordered by similarity - New case (fred case).

| Manager | BAM | Tolerance | | | | | | | | | bandwidth (Mb/s) | | | Measurements | | | | | | | | | | | | Solution | Similarity |
|---|---|---|---|---|---|---|---|---|---|---|---|---|---|---|---|---|---|---|---|---|---|---|---|---|---|---|---|
| | | Blocking | | | Preemption | | | Devolution | | | | | | Utilização | | | Blocking | | | Preemption | | | Devolution | | | | |
| | | TC0 | TC1 | TC2 | TC0 | TC1 | TC2 | TC0 | TC1 | TC2 | BC0 | BC1 | BC2 | TC0 | TC1 | TC2 | TC0 | TC1 | TC2 | TC0 | TC1 | TC2 | TC0 | TC1 | TC2 | | |
| **Fred** | **MAM** | **80** | **70** | **80** | **80** | **80** | **0** | **0** | **80** | **80** | **250** | **200** | **1000** | **10** | **20** | **30** | **0** | **0** | **0** | **0** | **0** | **0** | **0** | **0** | **0** | | |
| Carlos | MAM | 70 | 65 | 60 | 80 | 70 | 0 | 0 | 70 | 80 | 256 | 512 | 1024 | 10 | 15 | 20 | 20 | 25 | 30 | 0 | 0 | 0 | 0 | 0 | 0 | AllocCTS | 88% |
| Carlos | MAM | 70 | 65 | 60 | 80 | 70 | 0 | 0 | 70 | 80 | 128 | 256 | 512 | 10 | 15 | 20 | 20 | 25 | 30 | 0 | 0 | 0 | 0 | 0 | 0 | AllocCTS | 88% |
| Lucas | MAM | 65 | 60 | 70 | 70 | 60 | 0 | 0 | 60 | 70 | 256 | 512 | 1024 | 10 | 15 | 20 | 20 | 25 | 30 | 0 | 0 | 0 | 0 | 0 | 0 | AllocCTS | 87% |
| Lucas | MAM | 65 | 60 | 70 | 70 | 60 | 0 | 0 | 60 | 70 | 128 | 256 | 512 | 10 | 15 | 20 | 20 | 25 | 30 | 0 | 0 | 0 | 0 | 0 | 0 | AllocCTS | 87% |
| Marcos | MAM | 60 | 50 | 40 | 90 | 80 | 0 | 0 | 80 | 90 | 256 | 512 | 1024 | 10 | 15 | 20 | 20 | 25 | 30 | 0 | 0 | 0 | 0 | 0 | 0 | AllocCTS | 87% |
| Marcos | MAM | 60 | 50 | 40 | 90 | 80 | 0 | 0 | 80 | 90 | 128 | 256 | 512 | 10 | 15 | 20 | 20 | 25 | 30 | 0 | 0 | 0 | 0 | 0 | 0 | AllocCTS | 87% |
| Carlos | MAM | 70 | 65 | 60 | 80 | 70 | 0 | 0 | 70 | 80 | 256 | 512 | 1024 | 95 | 90 | 99 | 90 | 95 | 99 | 0 | 0 | 0 | 0 | 0 | 0 | MAN | 80% |
| Carlos | MAM | 70 | 65 | 60 | 80 | 70 | 0 | 0 | 70 | 80 | 128 | 256 | 512 | 95 | 90 | 99 | 90 | 95 | 99 | 0 | 0 | 0 | 0 | 0 | 0 | MAN | 80% |
| Lucas | MAM | 65 | 60 | 70 | 70 | 60 | 0 | 0 | 60 | 70 | 256 | 512 | 1024 | 95 | 90 | 99 | 90 | 95 | 99 | 0 | 0 | 0 | 0 | 0 | 0 | MAN | 78% |
| Lucas | MAM | 65 | 60 | 70 | 70 | 60 | 0 | 0 | 60 | 70 | 128 | 256 | 512 | 95 | 90 | 99 | 90 | 95 | 99 | 0 | 0 | 0 | 0 | 0 | 0 | MAN | 78% |
| Marcos | MAM | 60 | 50 | 40 | 90 | 80 | 0 | 0 | 80 | 90 | 256 | 512 | 1024 | 95 | 90 | 99 | 90 | 95 | 99 | 0 | 0 | 0 | 0 | 0 | 0 | MAN | 78% |
| Marcos | MAM | 60 | 50 | 40 | 90 | 80 | 0 | 0 | 80 | 90 | 128 | 256 | 512 | 95 | 90 | 99 | 90 | 95 | 99 | 0 | 0 | 0 | 0 | 0 | 0 | MAN | 78% |
| Carlos | RDM | 70 | 65 | 60 | 80 | 70 | 0 | 0 | 70 | 80 | 256 | 512 | 1024 | 10 | 15 | 20 | 20 | 25 | 30 | 0 | 0 | 0 | 0 | 0 | 0 | AllocCTS | 63% |
| Carlos | RDM | 70 | 65 | 60 | 80 | 70 | 0 | 0 | 70 | 80 | 128 | 256 | 512 | 10 | 15 | 20 | 20 | 25 | 30 | 0 | 0 | 0 | 0 | 0 | 0 | AllocCTS | 63% |
| Carlos | AllocCTS | 70 | 65 | 60 | 80 | 70 | 0 | 0 | 70 | 80 | 256 | 512 | 1024 | 10 | 15 | 20 | 20 | 25 | 30 | 0 | 0 | 0 | 0 | 0 | 0 | AllocCTS | 63% |
| Carlos | AllocCTS | 70 | 65 | 60 | 80 | 70 | 0 | 0 | 70 | 80 | 128 | 256 | 512 | 10 | 15 | 20 | 20 | 25 | 30 | 0 | 0 | 0 | 0 | 0 | 0 | AllocCTS | 63% |
| Lucas | RDM | 65 | 60 | 70 | 70 | 60 | 0 | 0 | 60 | 70 | 256 | 512 | 1024 | 10 | 15 | 20 | 20 | 25 | 30 | 0 | 0 | 0 | 0 | 0 | 0 | AllocCTS | 62% |
| Lucas | RDM | 65 | 60 | 70 | 70 | 60 | 0 | 0 | 60 | 70 | 128 | 256 | 512 | 10 | 15 | 20 | 20 | 25 | 30 | 0 | 0 | 0 | 0 | 0 | 0 | AllocCTS | 62% |
| Lucas | AllocCTS | 65 | 60 | 70 | 70 | 60 | 0 | 0 | 60 | 70 | 256 | 512 | 1024 | 10 | 15 | 20 | 20 | 25 | 30 | 0 | 0 | 0 | 0 | 0 | 0 | AllocCTS | 62% |
| Lucas | AllocCTS | 65 | 60 | 70 | 70 | 60 | 0 | 0 | 60 | 70 | 128 | 256 | 512 | 10 | 15 | 20 | 20 | 25 | 30 | 0 | 0 | 0 | 0 | 0 | 0 | AllocCTS | 62% |
| Marcos | RDM | 60 | 50 | 40 | 90 | 80 | 0 | 0 | 80 | 90 | 256 | 512 | 1024 | 10 | 15 | 20 | 20 | 25 | 30 | 0 | 0 | 0 | 0 | 0 | 0 | AllocCTS | 62% |
| Marcos | RDM | 60 | 50 | 40 | 90 | 80 | 0 | 0 | 80 | 90 | 128 | 256 | 512 | 10 | 15 | 20 | 20 | 25 | 30 | 0 | 0 | 0 | 0 | 0 | 0 | AllocCTS | 62% |
| Marcos | AllocCTS | 60 | 50 | 40 | 90 | 80 | 0 | 0 | 80 | 90 | 256 | 512 | 1024 | 10 | 15 | 20 | 20 | 25 | 30 | 0 | 0 | 0 | 0 | 0 | 0 | AllocCTS | 62% |
| Marcos | AllocCTS | 60 | 50 | 40 | 90 | 80 | 0 | 0 | 80 | 90 | 128 | 256 | 512 | 10 | 15 | 20 | 20 | 25 | 30 | 0 | 0 | 0 | 0 | 0 | 0 | AllocCTS | 62% |
| Carlos | RDM | 70 | 65 | 60 | 80 | 70 | 0 | 0 | 70 | 80 | 256 | 512 | 1024 | 95 | 90 | 99 | 0 | 0 | 0 | 80 | 90 | 0 | 0 | 0 | 0 | MAN | 57% |
| Carlos | RDM | 70 | 65 | 60 | 80 | 70 | 0 | 0 | 70 | 80 | 128 | 256 | 512 | 95 | 90 | 99 | 0 | 0 | 0 | 80 | 90 | 0 | 0 | 0 | 0 | MAN | 57% |
| Carlos | AllocCTS | 70 | 65 | 60 | 80 | 70 | 0 | 0 | 70 | 80 | 256 | 512 | 1024 | 95 | 90 | 99 | 0 | 0 | 0 | 80 | 90 | 0 | 0 | 0 | 0 | MAN | 57% |
| Carlos | AllocCTS | 70 | 65 | 60 | 80 | 70 | 0 | 0 | 70 | 80 | 128 | 256 | 512 | 95 | 90 | 99 | 0 | 0 | 0 | 80 | 90 | 0 | 0 | 0 | 0 | MAN | 57% |
| Carlos | AllocCTS | 70 | 65 | 60 | 80 | 70 | 0 | 0 | 70 | 80 | 256 | 512 | 1024 | 95 | 90 | 99 | 0 | 0 | 0 | 0 | 0 | 0 | 0 | 90 | 80 | RDM | 57% |
| Carlos | AllocCTS | 70 | 65 | 60 | 80 | 70 | 0 | 0 | 70 | 80 | 128 | 256 | 512 | 95 | 90 | 99 | 0 | 0 | 0 | 0 | 0 | 0 | 0 | 90 | 80 | RDM | 57% |
| Lucas | RDM | 65 | 60 | 70 | 70 | 60 | 0 | 0 | 60 | 70 | 256 | 512 | 1024 | 95 | 90 | 99 | 0 | 0 | 0 | 80 | 90 | 0 | 0 | 0 | 0 | MAN | 56% |
| Lucas | RDM | 65 | 60 | 70 | 70 | 60 | 0 | 0 | 60 | 70 | 128 | 256 | 512 | 95 | 90 | 99 | 0 | 0 | 0 | 80 | 90 | 0 | 0 | 0 | 0 | MAN | 56% |
| Lucas | AllocCTS | 65 | 60 | 70 | 70 | 60 | 0 | 0 | 60 | 70 | 256 | 512 | 1024 | 95 | 90 | 99 | 0 | 0 | 0 | 80 | 90 | 0 | 0 | 0 | 0 | MAN | 56% |
| Lucas | AllocCTS | 65 | 60 | 70 | 70 | 60 | 0 | 0 | 60 | 70 | 128 | 256 | 512 | 95 | 90 | 99 | 0 | 0 | 0 | 80 | 90 | 0 | 0 | 0 | 0 | MAN | 56% |
| Lucas | AllocCTS | 65 | 60 | 70 | 70 | 60 | 0 | 0 | 60 | 70 | 256 | 512 | 1024 | 95 | 90 | 99 | 0 | 0 | 0 | 0 | 0 | 0 | 0 | 90 | 80 | RDM | 56% |
| Lucas | AllocCTS | 65 | 60 | 70 | 70 | 60 | 0 | 0 | 60 | 70 | 128 | 256 | 512 | 95 | 90 | 99 | 0 | 0 | 0 | 0 | 0 | 0 | 0 | 90 | 80 | RDM | 56% |
| Marcos | RDM | 60 | 50 | 40 | 90 | 80 | 0 | 0 | 80 | 90 | 256 | 512 | 1024 | 95 | 90 | 99 | 0 | 0 | 0 | 80 | 90 | 0 | 0 | 0 | 0 | MAN | 56% |
| Marcos | RDM | 60 | 50 | 40 | 90 | 80 | 0 | 0 | 80 | 90 | 128 | 256 | 512 | 95 | 90 | 99 | 0 | 0 | 0 | 80 | 90 | 0 | 0 | 0 | 0 | MAN | 56% |

## VI. Conclusion

This paper focused on introducing, discussing and configuring in terms of a proof of concept the similarity functions that are essential in the CBR 4R retrieve phase and subsequent learning process.

The proof of concept presented shows that the similarity function used was capable to get similar results from the existing CBR database. As such, this suggests that, minimally, the CBR 4R approach supported by the similarity function will be able to estimate what is the most adequate BAM model for the problem-case that triggered the search and possibly learn about a new solution.

As such, the use of similarity functions with CBR techniques has proved to be potentially satisfactory for supporting BAM switching decisions that are mostly driven by the dynamics of input traffic profile.